\documentstyle[sprocl,epsf,fleqn]{article}

\def\be{\begin{equation}}
\def\ee{\end{equation}}
\def\bea{\begin{eqnarray}}
\def\eea{\end{eqnarray}}
\newcommand{\tr}{\rm Tr}
\newcommand{\nn}{\nonumber}

\begin{document}

\title{CONNECTING HIGGS AND CONFINEMENT PHASE OF THE 3D
SU(2) HIGGS MODEL}

\author{ O. PHILIPSEN }

\address{Institut f\"ur Theoretische Physik, Philosophenweg 16,\\
69120 Heidelberg, Germany}


\maketitle\abstracts{
The mass spectrum of the 3d SU(2) Higgs model for states with quantum numbers 
$0^{++}$, $2^{++}$ and $1^{--}$ as obtained from lattice simulations 
is presented in the Higgs and
confinement regions, both for a small and a large scalar coupling.
In the latter case it is possible to continuously connect the spectra
and observe the transition from confining to Higgs physics via 
decay of flux loops.} 
  
\section{Introduction}

The 3d SU(2) Higgs model has been subject of several Monte Carlo 
simulations in order to obtain a non-perturbative understanding of the 
electroweak phase transition in the framework of the dimensional
reduction program. In particular, it has been shown that the first order
phase transition terminates and
turns into a smooth crossover beyond some critical scalar coupling
\cite{laine}.
However, our understanding of the symmetric phase is still limited.
Lattice calculations of the mass spectrum from gauge-invariant 
operators lead to a picture of a confining symmetric phase with a
dense spectrum of bound states \cite{us}.
The connection between the mass spectrum and   
screening masses defined as poles of the
full propagators in a fixed gauge remains yet to be fully
clarified \cite{buc}.

The following is a summary of results 
obtained from lattice simulations 
in collaboration with M.~Teper and H.~Wittig \cite{us}.
We attempt to gain some insight in
the dynamics of confinement by exploiting  the analytic
connectedness of the phase diagram. The latter implies that all physical
states encountered in the Higgs region, which is
well understood perturbatively, may be mapped continuously onto their
counterparts in the non-perturbative confinement region.

The three-dimensional lattice action is
\begin{eqnarray} \label{actlat}
\lefteqn{S[U,\phi]=\beta_G\sum_p\left(1-\frac{1}{2}\tr U_p\right)
+\sum_x\Bigg\{-\beta_H\sum_{\mu=1}^3\frac{1}{2}
\tr\Big(\phi^{\dagger}_x U_{\mu x}
\phi_{x+\hat{\mu}}\Big)}\nn\\
&&+\frac{1}{2}\tr\Big(\phi^{\dagger}_x\phi_x\Big)
+\beta_R\left[\frac{1}{2}\tr\Big(\phi^{\dagger}_x\phi_x\Big)-1\right]^2
\Bigg\}\;,
\end{eqnarray}
where the continumm scalar to gauge coupling ratio
$\lambda_3/g_3^2 = (\beta_R\,\beta_G)/\beta_H^2$
fixes the Higgs to W boson mass ratio. 
All results have been obtained at $\beta_G=9$ where the masses
are known to be within the scaling region \cite{us}. 

\section{Operators and lattice techniques}

In the $0^{++}$,$2^{++}$-channel we consider gauge-invariant 
operators of various types,
such as ones containing only scalar fields, 
$R\sim \tr(\phi^{\dagger}_x\phi_x)$,
scalar fields and links,
$L_\mu\sim \tr(\phi^{\dagger}_x U_{\mu x} \phi_{x+\hat{\mu}})$,
and plaquettes $P\sim \tr (U_p)$ consisting of gauge degrees of freedom 
only.
In the $1^{--}$ channel we know only one gauge-invariant operator,
$V_\mu^a \sim \tr(\tau^a\phi^{\dagger}_x U_{\mu x}
\phi_{x+\hat{\mu}})$.
Another operator useful to clarify the properties of the regions in 
parameter space is the Polyakov loop operator,
$PL(i)\sim\tr \prod^L_{x_i=1} U_{i(x+n\hat{i})}$, whose expectation 
value vanishes in a confining theory. 
In the Higgs model one expects this never to be the
case as the string between fundamental charges eventually
breaks beyond some large separation due to matter pair creation.
In some of our simulations
in the symmetric phase, however, the measured VEV's are statistically
compatible with zero. This means that screening of the flux has not yet
set in for separations as large as the spatial lattice size, and
up to this distance the Wilson loop
still behaves according to the area law.
In these cases one may extract a volume-corrected
string tension from the correlations of Polyakov loops
according to \cite{for85}
\begin{equation} \label{sigma}
a^2\sigma=a^2\sigma_L+\frac{\pi}{6}\frac{1}{L^2};\quad
aM_{PL}(L)=a^2\sigma_LL\; ,
\end{equation}
where $L$ is the spatial length of the lattice.
 
In order to improve the projection properties of our operators
we ``smear" the fields by covariantly
connecting them with their neighbours \cite{us}. 
The excitation spectrum may be computed
by considering $N$ operators $\phi_i$ of different types and 
smearing levels for each quantum number and measure  
correlations between all of them.
This correlation matrix can then be diagonalised numerically 
\cite{us} resulting in a set of $N$ (approximate)
mass eigenstates of the Hamiltonian
$\Phi_i =  \sum_{k=1}^N a_{ik}\phi_k$, which are superpositions of the 
operators $\phi_i$ used in the simulation.
The coefficients $a_{ik}$
are useful in identifying the
contributions of the individual operators $\phi_i$ to the
mass eigenstates.

\section{The spectrum at small and large scalar coupling}

\begin{figure}[ht]
\begin{center}
\leavevmode
\epsfysize=135pt
\epsfbox[20 30 620 730]{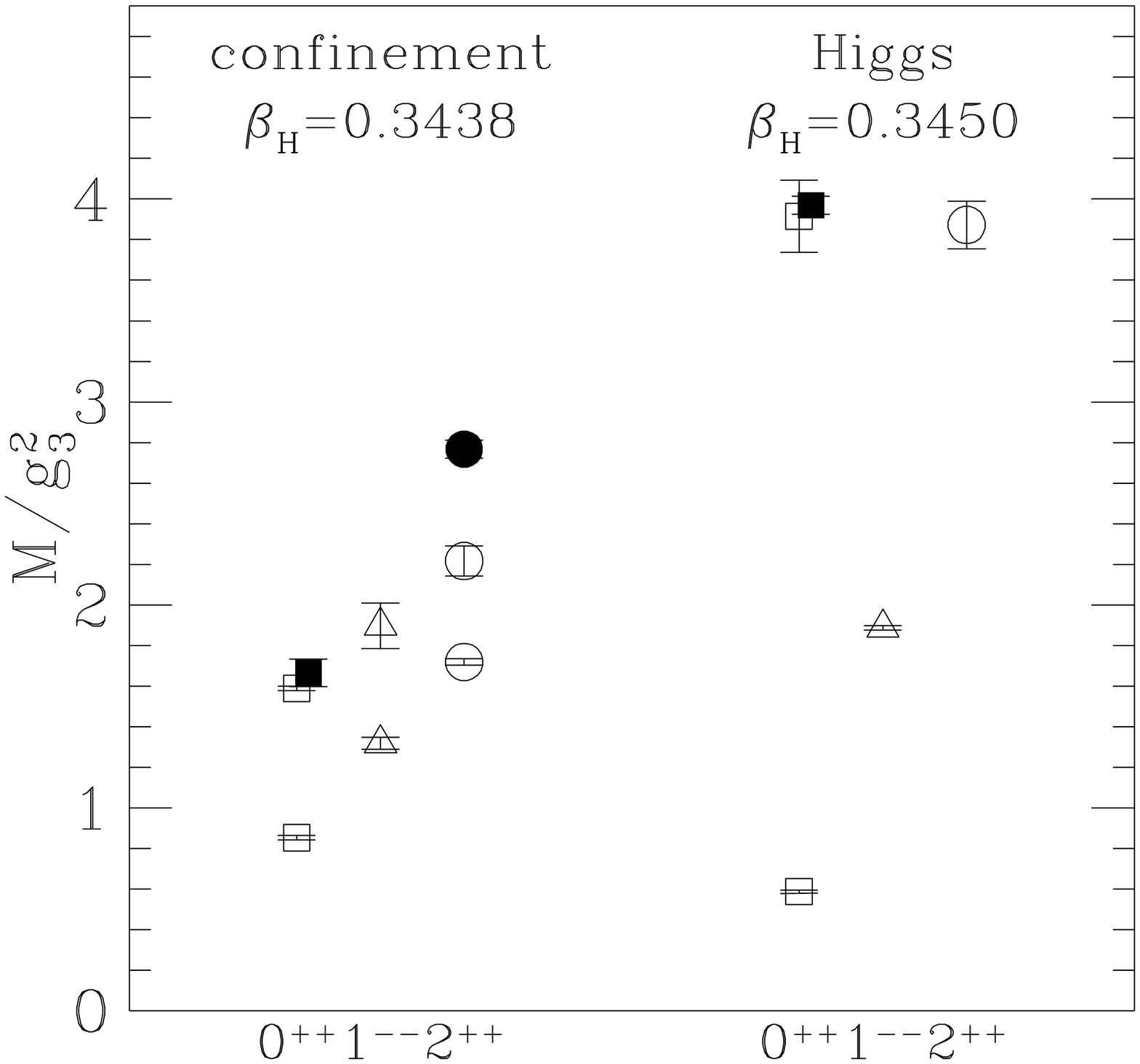}
\leavevmode
\epsfysize=135pt
\epsfbox[20 30 620 730]{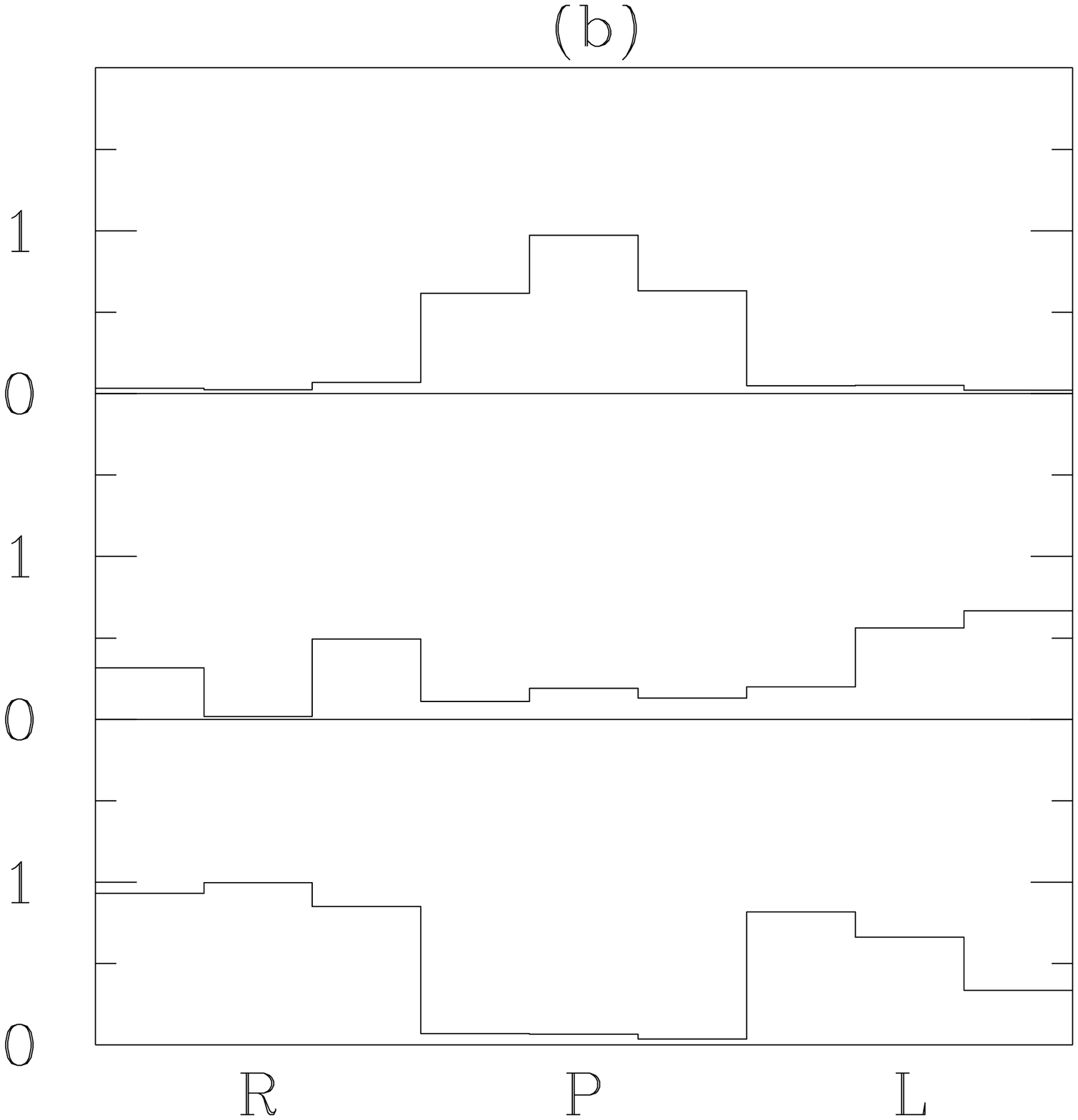}
\leavevmode
\epsfysize=135pt
\epsfbox[20 30 620 730]{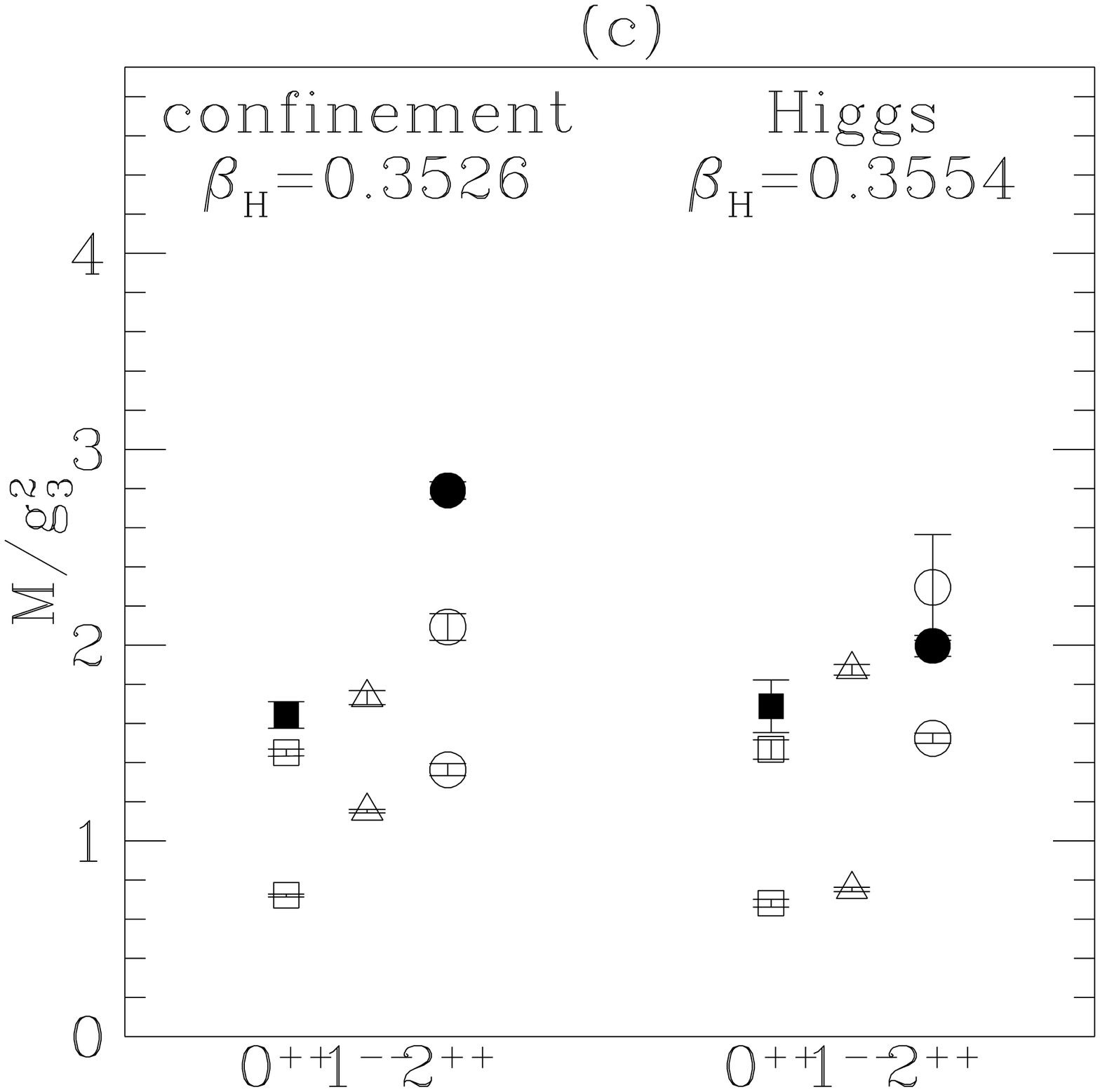}
\vspace{-1.0cm}
\end{center}
\caption[]{\it \label{spec}
The lowest states of the spectrum in the confinement (left)
and Higgs (right) regions for (a) $\lambda_3/g_3^2=0.0239$,
(c) $\lambda_3/g_3^2=0.274$. Full symbols denote pure gauge states.
(b) Coefficients $a_{ik}$ of the $0^{++}$ states on the confinement side of (a), $\phi_i\in \{R,P,L\}$.}
\end{figure}

The mass spectrum at small scalar coupling $\lambda_3/g_3^2=0.0239$
is shown in Fig.~\ref{spec} (a). 
In the Higgs phase there are the familiar Higgs and W bosons with a large gap
to higher excitations representing
scattering states with relative momentum.
In the symmetric phase, on the other hand, there is a dense spectrum of
bound states.
Their composition may be characterised by considering the contributions
$a_{ik}$ of the individual operators to each eigenstate, Fig.~\ref{spec} (b).
The pure gauge plaquette operators, $P$,
contribute very little to the
ground state and the first excited state.
However, the third $0^{++}$ state is composed almost entirely of them.
This suggests interpreting it as a
``W-ball", in analogy to the glueballs of pure gauge theory.
In the Higgs phase the plaquette projects onto a two-W scattering state,
in agreement with perturbation theory.
W-balls are also observed in the $2^{++}$-channel. 
The W-ball masses agree at the percent level with their pure gauge 
analogues \cite{for85}, 
indicating a fairly complete decoupling of the pure gauge sector. 
In Table 1 some properties of the Polyakov loop are listed.
In the confinement phase its VEV is compatible with zero and one may
extract a string tension which is about $97\%$ of the one of pure gauge
theory \cite{for85}. On the Higgs side it has a large VEV and no string
tension exists. Instead, a perturbative expansion of the correlator is
possible whose leading term corresponds to a two-W state. Indeed, we 
find the effective mass of the Polyakov loop correlator to be compatible
with twice the W-mass.
\begin{table}[ht]
\begin{center}
\label{pol}
\begin{tabular}{|l|r@{.}l|r@{.}l|}
\hline
  & \multicolumn{2}{c|}{conf.}
  & \multicolumn{2}{c|}{Higgs} \\ \hline
$\langle PL \rangle$ & 0&001(1)  & 6&535(6) \\ \hline
$aM_{PL}$            & 0&577(8)  & 1&8(1)  \\ \hline
$aM_W$               & 0&610(4)  & 0&836(3)\\ \hline
$a\sqrt{\sigma}$     & 0&1582(6) & \multicolumn{2}{c|}{--}  \\ \hline
\end{tabular}
\caption[]{Properties of the Polyakov loop, 
$\lambda_3/g_3^2=0.0239$.}
\end{center}
\end{table}

Now consider the large scalar coupling $\lambda_3/g_3^2=0.274$, where
there is no phase transition anymore separating the two regions.
The spectrum for this case is displayed in Fig.~\ref{spec} (c).
On the confinement side it qualitatively looks the same as before.
We even find the masses of the W-ball states to be numerically
compatible with those of the small scalar coupling case. This confirms
the decoupling of the pure gauge sector over a range of scalar couplings
of an order of magnitude!
On the Higgs side, a dramatic change has taken place,
with the spectrum now qualitatively resembling 
that on the confinement side. Accordingly, perturbation theory is not
applicable and one cannot identify the effective mass of the Polyakov
loop with a two-W state. 
Does the confining dynamics now extend to the
other side of the crossover? Considering again the Polyakov
loop, Fig.~\ref{m0} (a), 
we see that this is not the case. Whereas the usual ``order
parameter" $\langle R \rangle$ used to distinguish the phases only grows
about 30 \% from the confinement point to the Higgs point, 
the VEV of the Polyakov
loop shows a pronounced increase. 
Despite the smoothness of the crossover
a critical coupling $\beta_H^c$ may be defined
by the peak in the susceptibility of $R$,
$\chi=  \langle R^2 \rangle - \langle R \rangle^2 $, which separates
the Higgs and confinement regions, Fig.~\ref{m0} (b).
On the confinement side 
the effective mass of the Polyakov loop
increases linearly with the lattice size as required
to interpret it as a flux loop. 
On the Higgs side the VEV is large and no flux loops exist. 
We are thus led to conclude that the latter corresponds to
a Higgs regime with strong
scalar coupling. 

\section{Continuous connection of the spectrum and flux loop decay}

\begin{figure}[ht]
\begin{center}
\leavevmode
\epsfysize=135pt
\epsfbox[20 30 620 730]{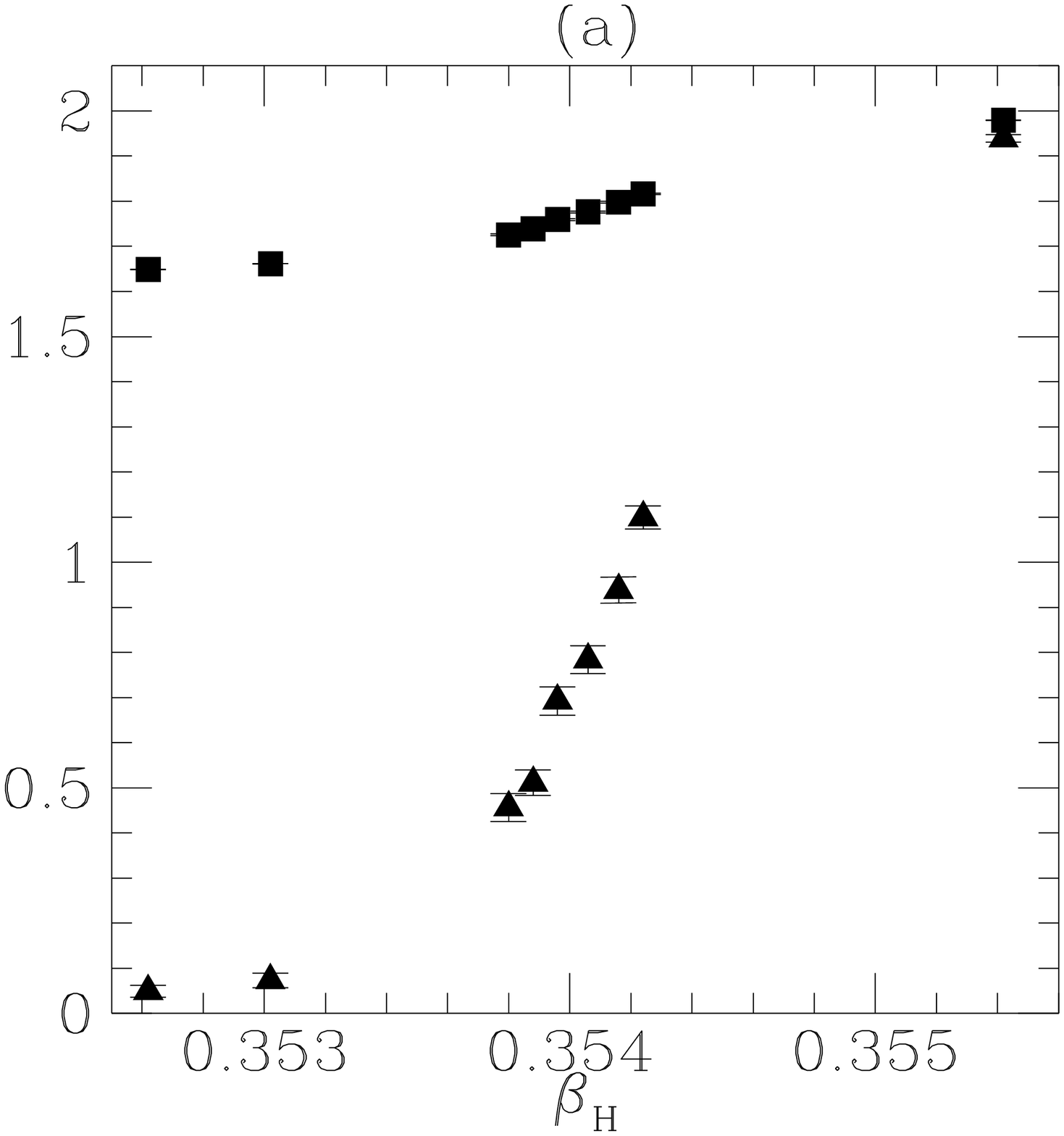}
\leavevmode
\epsfysize=135pt
\epsfbox[20 30 620 730]{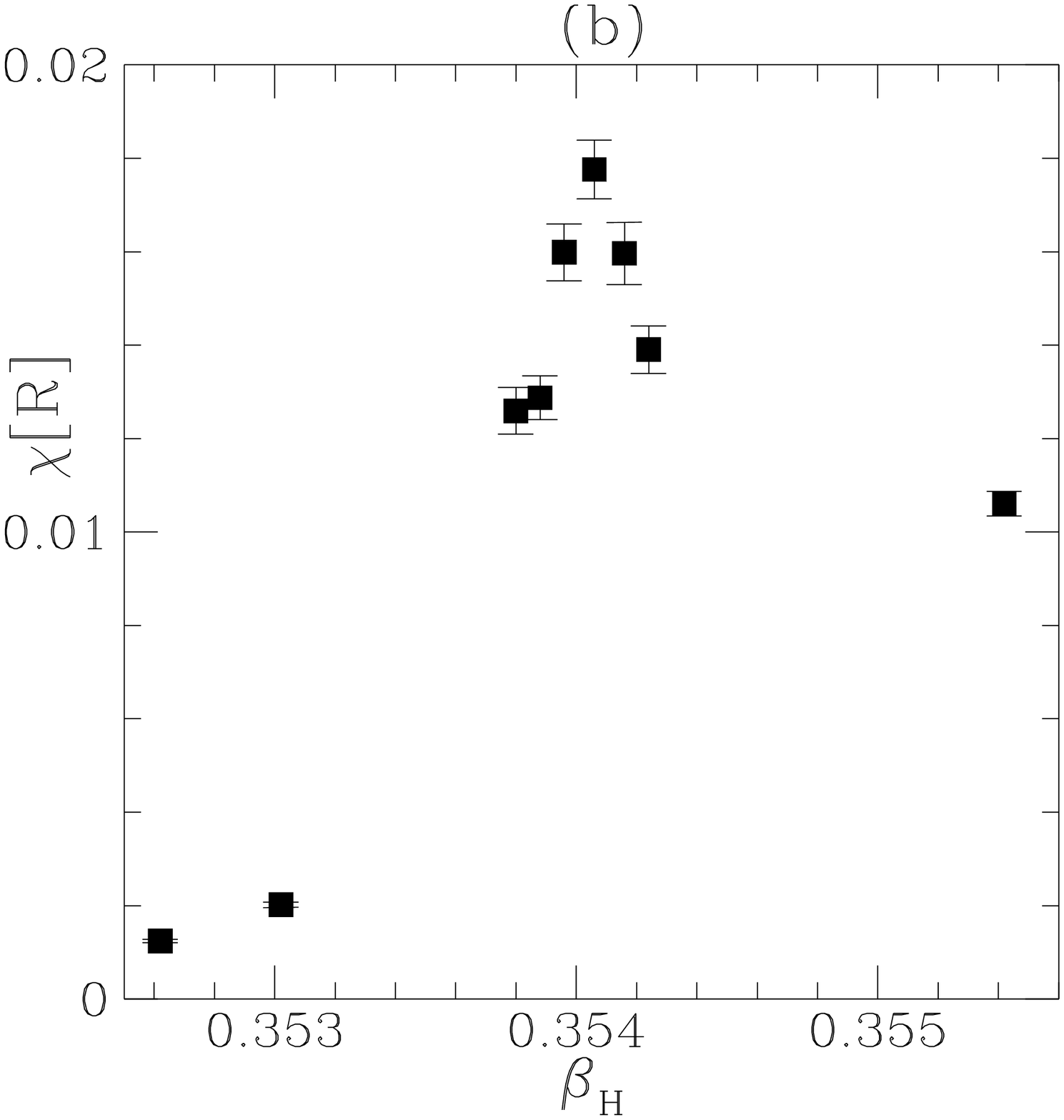}
\leavevmode
\epsfysize=135pt
\epsfbox[20 30 620 730]{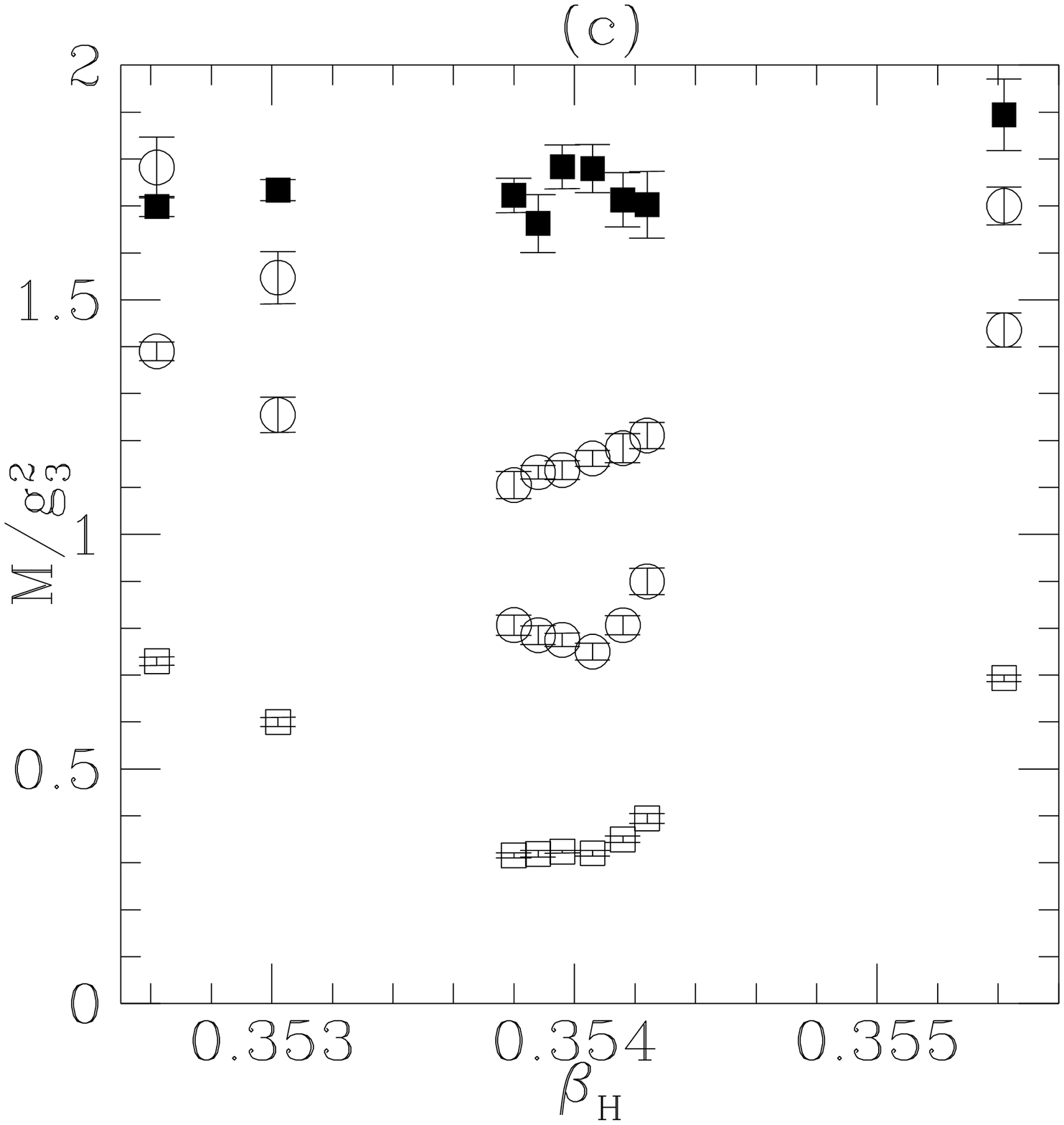}
\vspace{-1.0cm}
\end{center}
\caption[]{\it \label{m0}
(a) Squares: $\langle R \rangle$, triangles: $\langle PL \rangle$.
(b) The susceptibility $\chi=  \langle R^2 \rangle - \langle R \rangle^2 $.\\
(c) The $0^{++}$ states through the crossover. 
Full symbols denote the pure gauge state.}
\end{figure}

The four lowest $0^{++}$-states are connected through the crossover as 
shown in Fig.~\ref{m0} (c). Note that the ground state
dips but stays finite in accordance with the absence of a diverging
correlation length in a crossover.
In contrast, the mass of the W-ball is independent of $\beta_H$ until 
beyond 
the critical coupling, which is yet another manifestation of its
decoupling.

\begin{figure}[ht]
\begin{center}
\leavevmode
\epsfysize=135pt
\epsfbox[20 30 620 730]{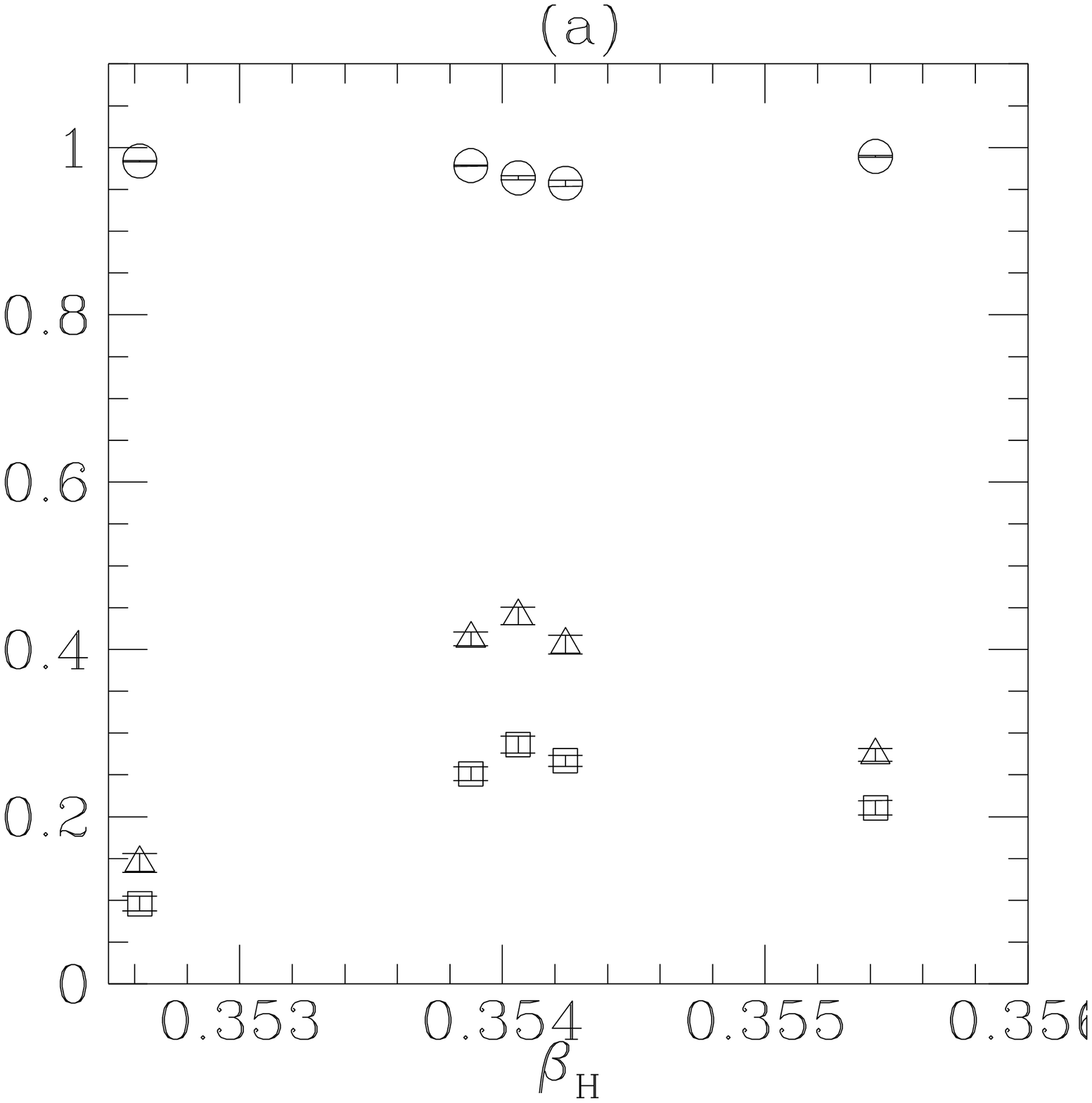}
\leavevmode
\epsfysize=135pt
\epsfbox[20 30 620 730]{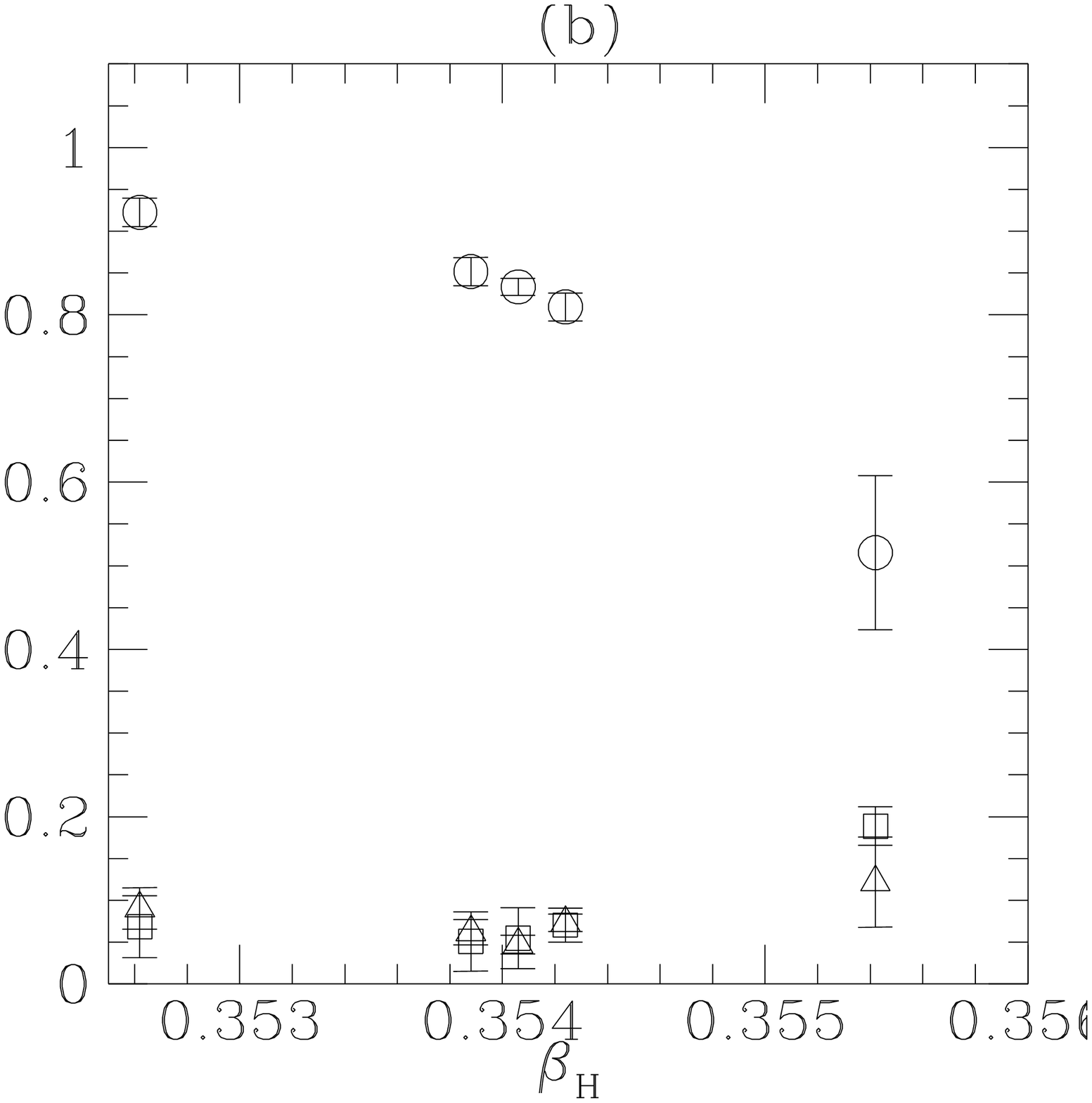}
\leavevmode
\epsfysize=135pt
\epsfbox[20 30 620 730]{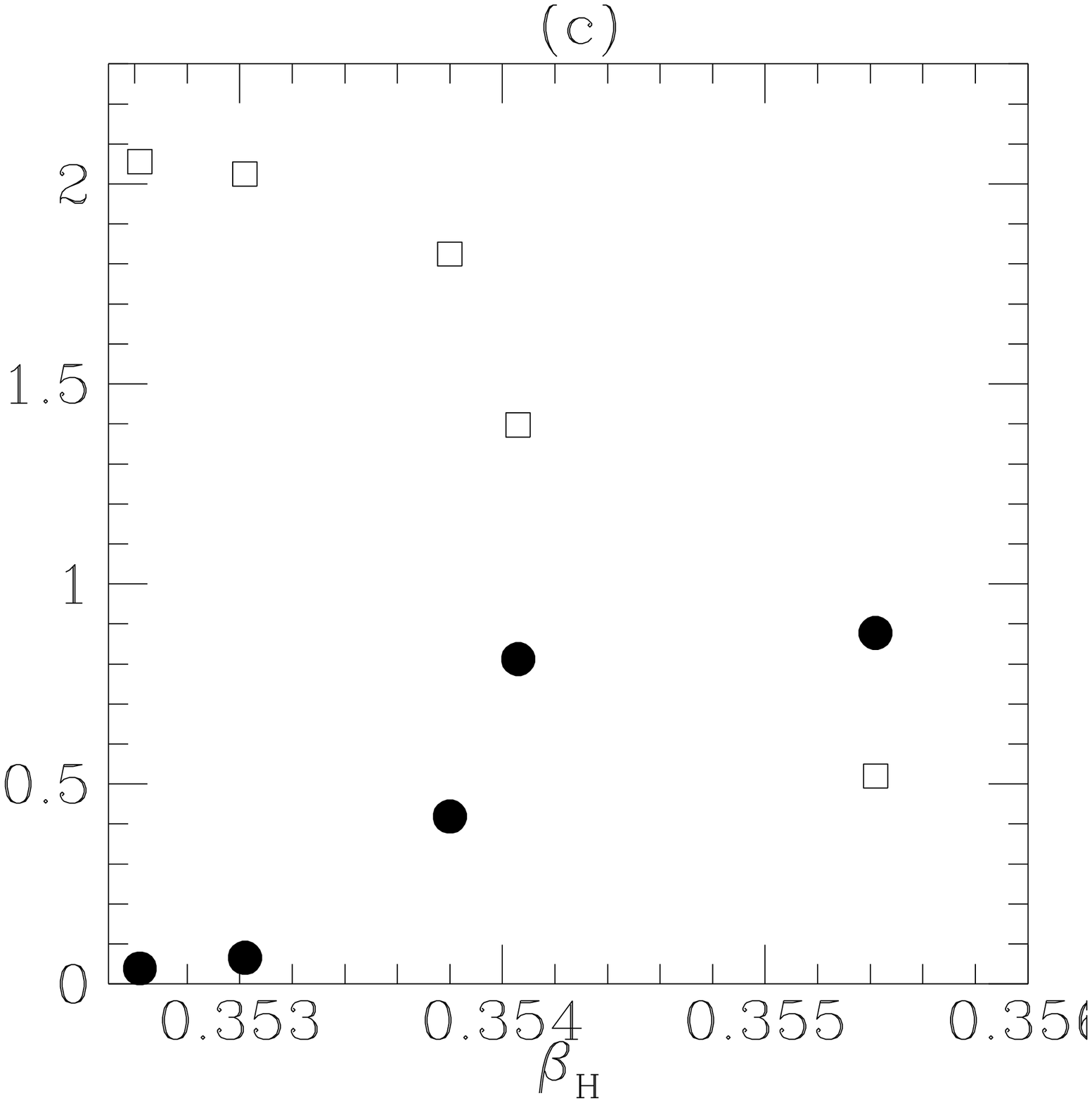}
\vspace{-1.0cm}
\end{center}
\caption[]{\it \label{gamma}
The $a_{ik}$ of the (a) $0^{++}$ ground state and 
(b) the $0^{++}$ W-ball. Circles denote $R/L$ operators, squares
$P$ and triangles $PL$. (c) Squares: $\langle E_F\rangle$,
full circles: $ \Gamma_F $.}
\end{figure}

In Fig.~\ref{gamma} (a) the coefficents of the different operator types in
the $0^{++}$ ground state are shown.
Throughout the crossover the dominating 
contribution comes from scalar operators. However, moving towards the
critical point, there is an increasing contribution from plaquettes and
Polyakov loops which decreases again on the Higgs side.
Due to decoupling, the converse is not true for the W-ball, Fig.~\ref{gamma} (b).
The growing overlap of the Polyakov loop with the scalar ground state
has a physical interpretation. Recall that the Polyakov loop
projects on a flux loop 
in the confinement phase which does not 
exist on the Higgs side. The overlap with scalar states
signals an increasing coupling between the flux loop and these states. 
A natural physical picture then is that the flux loop becomes
increasingly unstable and eventually decays, where the $0^{++}$ eigenstates
are some of the possible decay products.
Consequently the pole in the $PL$ correlator moves away from the
real axis. Its real part is given by the weighted 
sum of the energies of the decay products,
the imaginary part by its decay width.
To test this picture 
we define the ``effective flux loop energy" $E_F$ and the corresponding
``decay width" $\Gamma_F$ by
\begin{equation}
\langle E_F \rangle \equiv \frac{\sum_i \mid a_{PL,i} \mid ^2 M_i}
{\sum_i \mid a_{PL,i} \mid ^2} \;, \qquad
\Gamma_F^2=\langle E_F^2 \rangle - \langle E_F \rangle ^2\;.
\end{equation}
Numerical results for these are shown in Fig.~\ref{gamma}
(c). The decay width is close to zero on the confinement side where
flux loops are stable, and then increases towards the critical
coupling maintaining a high value on the Higgs side. 
We conclude that the picture of a decaying flux 
loop is qualitatively confirmed.

\end{document}